\input{aipcheck}

\documentclass[
    ,final            
  ]
  {aipproc}
\usepackage{amssymb}
\usepackage{epsfig}
\layoutstyle{8x11single}

\begin{document}
\newcommand{\A}{{\mathcal{A}}}
\newcommand{\dA}{\delta{\mathcal{A}}}
\newcommand{\Od}{{\cal O}}
\title{Electromagnetic nature of dark energy}


\author{Jose Beltr\'an Jim\'enez and Antonio L. Maroto}{
  address={Departamento de  F\'{\i}sica Te\'orica,
 Universidad Complutense de
  Madrid, 28040 Madrid, Spain}
}

\begin{abstract}
 Out of the four components of the electromagnetic field,
Maxwell's theory  only contains two physical degrees of 
freedom.  However, in an expanding universe, consistently 
eliminating one of the
 "unphysical" states in 
the covariant (Gupta-Bleuler) formalism turns out to be 
difficult to realize. In this work we explore the 
possibility of quantization without subsidiary conditions. This
implies that the theory would contain a third physical state.  
 The presence of such a new (temporal) electromagnetic mode on 
cosmological scales is shown to generate an effective cosmological 
constant which can account for the accelerated expansion 
of the universe. This new polarization state is completely
decoupled from charged matter, but can be excited 
gravitationally. In fact,  primordial 
electromagnetic quantum fluctuations produced during 
electroweak scale inflation could naturally explain the 
presence of this mode and also the measured value of the 
cosmological constant. The theory is compatible with all the 
local gravity tests, it is free from classical or quantum 
instabilities and reduces to standard QED in the flat space-time limit. 
Thus we see that, not only the true nature of 
dark energy can be established without resorting to new 
physics, but also the value of the cosmological constant 
finds a natural explanation in the context of standard 
inflationary cosmology. Possible signals, discriminating this 
model from LCDM, are also discussed.  

\end{abstract}

\maketitle


\section{What is the nature of dark energy?}
Although the standard cosmological constant provides a simple
and accurate phenomenological description of the present phase of 
accelerated expansion of the universe \cite{acc}, it is well known that it
suffers from an important naturalness problem. Indeed, the scale of the
cosmological constant around $10^{-3}$ eV is many orders of magnitude 
smaller than the fundamental gravity scale given by the Planck constant
which is around $10^{19}$ GeV.

Alternative models based on new physics have been proposed, 
however they 
are generally plagued by:
\begin{itemize}
\item classical or quantum instabilities
\item fine tuning problems
\item inconsistencies with local gravity constraints
\end{itemize}

Also, the possibility that the accelerated expansion could be signaling the
breakdown of General Relativity on large 
scales has been also proposed, and several modified gravity theories have been
considered \cite{modified,vector}, although, in general, they also suffer from the same kind of
problems described above.   

However, apart from gravity there is another long-range interaction in nature
which could be relevant on cosmological scales which is nothing but electromagnetism.
It is usually assumed that due to the electric neutrality of the universe on 
large scales, the only relevant interaction in cosmology should be gravitation. However,
 the role played by electromagnetic astrophysical and cosmological fields 
is still far from clear, being the existence of cosmic magnetic fields the most evident 
example \cite{Grasso}. As a matter of fact, and as for any other interaction in nature, 
electromagnetism has only been tested in a limited range of 
energies or distances, which are roughly  below 1.3 A.U. \cite{Nieto} and, as a consequence,
 no  experimental evidence is available for electromagnetic fields whose 
wavelenghts are comparable to the present Hubble radius.

In this work, we consider  (quantum) electromagnetic fields in an 
expanding universe. Unlike previous works, we will not follow the
standard Coulomb gauge approach, but instead we will concentrate
on the covariant quantization method. This 
method is found to exhibit certain difficulties when trying to impose the quantum Lorenz
condition on cosmological scales \cite{Parker}. In order to overcome these problems, 
and instead of introducing ghosts fields,  we explore the possibility of
quantizing without imposing any subsidiary condition \cite{EM,sector}. 
This implies that the
electromagnetic field would contain an additional (scalar)
polarization, similar to that generated in the massless limit
of massive electrodynamics \cite{Deser}. Interestingly, the energy density of quantum
fluctuations of this new electromagnetic state generated during
inflation gets frozen on cosmological scales, giving rise to an 
effective cosmological constant \cite{EM}.

\section{Electromagnetic quantization in flat space-time}

Let us start by briefly reviewing electromagnetic
quantization in Minkowski space-time \cite{Itzykson}
since this will be useful in the rest of the work.
The action
of the theory reads:
\begin{eqnarray}
S=\int d^4x \left(-\frac{1}{4}F_{\mu\nu}F^{\mu\nu}+A_\mu J^\mu\right)
 \label{action}
\end{eqnarray}
where $J_\mu$ is a conserved current. This action is invariant under
gauge tranformations $A_\mu\rightarrow A_\mu+\partial_\mu \Lambda$
with $\Lambda$ an arbitrary function of space-time coordinates.
At the classical level, this action gives rise to the well-known
Maxwell's equations:
\begin{eqnarray}
\partial_\nu F^{\mu\nu}=J^\mu.
\label{Maxwell}
\end{eqnarray}
However, when trying to quantize the theory, several problems
arise related to the redundancy in the description  
due to the gauge invariance. Thus, in particular, we find that it is 
not possible to construct a propagator for the $A_\mu$ field
and  also that "unphysical" polarizations of the 
photon field are present.  Two different approaches are usually followed in order
to avoid these difficulties. In the first one, which is the basis
of the Coulomb gauge quantization, the gauge invariance of the
action (\ref{action}) is used to eliminate the "unphysical"
degrees of freedom. With that purpose the (Lorenz) condition
$\partial_\mu A^\mu=0$ is imposed by means of a suitable gauge
transformation. Thus, the equations of motion reduce to:
\begin{eqnarray}
\Box A_\mu=J_\mu.\label{eqLo}
\end{eqnarray}
The Lorenz condition does not fix completely the gauge freedom,
still it is possible to perform residual gauge transformations
$A_\mu\rightarrow A_\mu+\partial_\mu \theta$, provided $\Box
\theta=0$. Using this residual symmetry and taking into account
the form of equations (\ref{eqLo}), it is possible to eliminate
one additional component of the $A_\mu$ field in the
asymptotically free regions
 (typically $A_0$) which means $\vec \nabla \cdot \vec A=0$,
so that finally the temporal and longitudinal
photons are removed and we are left with the two
transverse polarizations of the massless free photon, which are the
only modes (with positive energies)
which are quantized in this formalism.

The second approach is  the basis of the covariant (Gupta-Bleuler)
and path-integral formalisms. The starting point is a modification of the
action
in (\ref{action}), namely:
\begin{eqnarray}
S=\int d^4x \left(-\frac{1}{4}F_{\mu\nu}F^{\mu\nu}+\frac{\xi}{2}
(\partial_\mu A^\mu)^2+ A_\mu J^\mu\right).
 \label{actionGB}
\end{eqnarray}
This action is no longer invariant under general gauge transformations, but
only under residual ones.
The equations of motion obtained from this action now read:
\begin{eqnarray}
\partial_\nu F^{\mu\nu}+\xi\partial^\mu(\partial_\nu
A^\nu)=J^\mu.
\label{fieldeq}
\end{eqnarray}
In order to recover Maxwell's equation, the Lorenz condition must
be imposed so that the $\xi$ term disappears. At the classical
level this can be achieved by means of appropriate boundary
conditions on the field. Indeed, taking the four-divergence of the
above equation, we find:
\begin{eqnarray}
\Box(\partial_\nu A^\nu)=0
\end{eqnarray}
where we have made use of current conservation. This means that
the field  $\partial_\nu A^\nu$ evolves as a free scalar field, so
that if it vanishes for large $\vert t \vert$, it will vanish at 
all times. At the quantum level, the Lorenz condition cannot be
imposed as an operator identity, but only in the weak sense
$\partial_\nu A^{\nu \,(+)}\vert \phi\rangle=0$, where $(+)$
denotes the positive frequency part of the operator and $\vert
\phi\rangle$ is a physical state. This condition is
equivalent to imposing $[{\bf a}_0(\vec k) +{\bf a}_\parallel(\vec
k)] |\phi\rangle=0$, with ${\bf a}_0$ and ${\bf a}_\parallel$ the
annihilation operators corresponding to temporal and longitudinal
electromagnetic states. Thus, in the covariant formalism, the
physical states contain the same number of temporal and
longitudinal photons, so that their energy densities, having
opposite signs, cancel each other.

Thus we see that also in this case, the Lorenz condition seems to
be essential in order to recover standard Maxwell's equations and
get rid of the negative energy states.

\section{Covariant quantization in an expanding universe}

So far we have only considered Maxwell's theory in flat
space-time, however when we move to a curved background, and in
particular to an expanding universe, then consistently imposing
the Lorenz condition in the covariant formalism turns out to be
difficult to realize. Indeed, let us consider the curved
space-time version of action (\ref{actionGB}):
\begin{eqnarray}
S=\int d^4x
\sqrt{g}\left[-\frac{1}{4}F_{\mu\nu}F^{\mu\nu}+\frac{\xi}{2}
(\nabla_\mu A^\mu)^2+ A_\mu J^\mu\right]
 \label{actionF}
\end{eqnarray}
Now the modified Maxwell's equations read:
\begin{eqnarray}
\nabla_\nu F^{\mu\nu}+\xi\nabla^\mu(\nabla_\nu A^\nu)=J^\mu
\label{EMeqexp}
\end{eqnarray}
and taking again the four divergence, we get:
\begin{eqnarray}
\Box(\nabla_\nu A^\nu)=0\label{minimal}
\end{eqnarray}
We see that once again  $\nabla_\nu A^\nu$  behaves as a scalar
field which is decoupled from the conserved electromagnetic
currents, but it is non-conformally coupled to gravity. This means
that, unlike the flat space-time case,  this field can be excited
from quantum vacuum fluctuations by the expanding background in a
completely analogous way to the inflaton fluctuations during
inflation. Thus this poses the question of the validity of the
Lorenz condition at all times.

In order to illustrate this effect, we will present a toy example.
Let us consider quantization in the absence of currents, in a
spatially flat expanding background, whose metric is written in
conformal time as $ds^2=a(\eta)^2(d\eta^2-d\vec x^2)$ with
$a(\eta)=2+\tanh(\eta/\eta_0)$ where $\eta_0$ is constant. This
metric possesses two asymptotically Minkowskian regions in the
remote past and far future. We  solve the coupled system of
equations (\ref{EMeqexp}) for the corresponding Fourier modes,
which are defined as $\A_\mu(\eta,\vec x)= \int d^3k\A_{\mu
\vec k}(\eta) e^{i\vec k \vec x}$. Thus, for a given mode $\vec k$, the
$\A_\mu$ field is  decomposed into temporal,  longitudinal and
transverse components. The corresponding equations read:
\begin{eqnarray}
\A_{0k}''&-&\left[\frac{k^2}{\xi}-2{\mathcal{H}}'
+4{\mathcal{H}}^2\right]\A_{0k}
-2ik\left[\frac{1+\xi}{2\xi}\A_{\parallel k}'
-{\mathcal{H}}\A_{\parallel k}\right]=0 \label{modes}\nonumber\\
\A_{\parallel k}''&-&k^2\xi\A_{\parallel
k}-2ik\xi\left[\frac{1+\xi}{2\xi}\A_{0k}'
+{\mathcal{H}}\A_{0k}\right]=0\nonumber\\
\vec{\A}_{\perp k}''&+&k^2\vec{\A}_{\perp k }=0
\end{eqnarray}
with ${\cal H}=a'/a$ and $k=\vert \vec k\vert$. We see that the transverse 
modes are decoupled
from the background, whereas the temporal and longitudinal ones are
non-trivially coupled to each other and to gravity. Let us prepare our
system  in an initial  state $\vert \phi\rangle$ belonging
to the physical Hilbert space,
i.e. satisfying $\partial_\nu \A^{\nu \,(+)}_{in}\vert \phi\rangle=0$
in the initial flat region.
Because of the expansion
of the universe, the positive frequency modes in the $in$
region with a given temporal or longitudinal polarization $\lambda$
will become a linear superposition of positive and
negative frequency modes in the $out$ region and
with different polarizations $\lambda'$ (in this section we will work in
the Feynman gauge $\xi=-1$), thus we have:
\begin{eqnarray}
\A_{\mu \vec k}^{\lambda \; (in)}=\sum_{\lambda'=0,\parallel}
\left[\alpha_{\lambda\lambda'}(\vec k) \A_{\mu \,\vec k}^{\lambda' \;
(out)}+\beta_{\lambda\lambda'}(\vec k) \overline{\A_{\mu
\,-\vec k}^{\lambda' \; (out)}}\,\right]
\end{eqnarray}
or in terms of creation and annihilation operators:
\begin{equation}
{\bf a}_{\lambda'}^{(out)}(\vec k)=
\sum_{\lambda=0,\parallel}\left[\alpha_{\lambda\lambda'}(\vec k)
{\bf a}_{\lambda}^{(in)}(\vec
k)+\overline{\beta_{\lambda\lambda'}(\vec k)} {\bf
a}_{\lambda}^{(in)\dagger}(-\vec k)\right]
\end{equation}
with $\lambda, \lambda'=0,\parallel$ and
where $\alpha_{\lambda\lambda'}$ and $\beta_{\lambda\lambda'}$
are the so-called Bogolyubov coefficients (see \cite{Birrell} 
for a detailed discussion), which are  normalized in our case according to:
\begin{eqnarray}
\sum_{\rho,\rho'=0,\parallel}(\alpha_{\lambda \rho}\,
\overline{\alpha_{\lambda'\rho'}}\,
\eta_{\rho \rho'}
-\beta_{\lambda \rho}\,\overline{\beta_{\lambda'\rho'}}\,\eta_{\rho\rho'})=
\eta_{\lambda\lambda'}
\end{eqnarray}
with $\eta_{\lambda\lambda'}=diag(-1,1)$ with $\lambda,\lambda'=0,\parallel$.
Notice that the normalization is different from the standard one
\cite{Birrell}, because of the presence of  negative norm states.

\begin{figure}
  \includegraphics[height=.3\textheight]{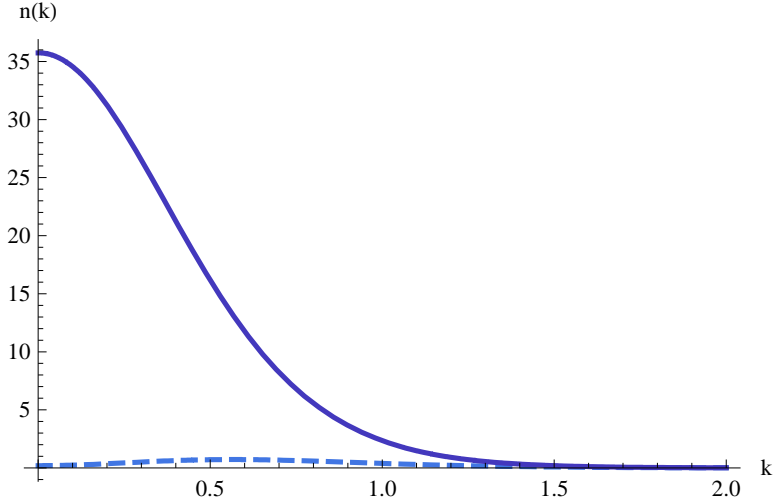}
  \caption{Occupation numbers for temporal (continuous
line) and longitudinal (dashed line) photons 
in the $out$ region vs.
$k$ in $\eta_0^{-1}$ units.}
\end{figure}

Thus, the system will end up in a final state which no longer
satisfies the weak Lorenz condition i.e. in the {\it out} region
$\partial_\nu \A^{\nu \,(+)}_{out}\vert \phi\rangle\neq 0$. This
is shown in Fig. 1, where we have computed  the final number of
temporal and longitudinal photons
$n_{\lambda'}^{out}(k)=\sum_{\lambda}\vert\beta_{\lambda\lambda'}(\vec
k) \vert^2 $, starting from an initial vacuum state with
$n_0^{in}(k)=n_\parallel^{in}(k)=0$. We see that, as commented
above, in the final region $n_0^{out}(k)\neq n_\parallel^{out}(k)$
and the state no longer satisfies the Lorenz condition. Notice
that the failure comes essentially from large scales, i.e. super-Hubble scales
with $k\eta_0\ll 1$, since on small (sub-Hubble) scales with $k\eta_0\gg 1$, 
the Lorenz condition is restored. This can be easily  interpreted from the fact
that on small scales the geometry can be considered as essentially
Minkowskian. 

In order to overcome this problem, it is possible to impose 
a more stringent gauge-fixing condition. Indeed, we have shown above that 
in a space-time configuration with asymptotic flat regions, an initial
state satisfying the weak Lorenz condition does not necessarily satisfy it
at a later time. However, it would be possible (see \cite{Pfenning}) to define 
the physical states $\vert \phi\rangle$ as those such that 
$\nabla_\mu A^{\mu (+)}\vert \phi\rangle =0$, 
$\forall \eta$. Although this is a perfectly consistent solution, notice that 
the separation in positive and negative frequency parts depends on 
the space-time geometry and therefore, the determination of the physical states 
requires a previous knowledge of  the future evolution of the universe. 

Another possible way out would be to modify the standard Gupta-Bleuler formalism
by including ghosts fields as done in non-abelian gauge theories \cite{Kugo}.
With that purpose, the action of the theory (\ref{actionF}) can be modified by including
the ghost term (see \cite{Adler}):
\begin{eqnarray}
S_{g}=\int d^4x\sqrt{g}\, g^{\mu\nu}\,\partial_\mu \bar c\, \partial_\nu c
\label{ghost}
\end{eqnarray} 
where $c$ are the complex scalar ghost fields. It is a well-known result 
\cite{Adler,BO} that by choosing appropriate boundary conditions  for the
electromagnetic and ghosts Green's functions, it is possible to get  
$\langle \phi\vert T_{\mu\nu}^\xi+T_{\mu\nu}^g\vert \phi\rangle=0$, where
$T_{\mu\nu}^\xi$ and $T_{\mu\nu}^g$ denote the contribution to the  energy-momentum
tensor from the $\xi$ term in (\ref{actionF}) and from the ghost term (\ref{ghost})
respectively.
Notice that a choice of boundary conditions in the Green's functions 
corresponds to a choice of  vacuum state. Therefore, also in this case  
an a priori knowledge of the future behaviour of the universe expansion 
is required in order to determine the physical states.

Although these are valid quanization procedures, 
in this work we follow a different approach in order to deal with the 
difficulties found in the Gupta-Bleuler formalism  and  we will explore
the possibility of quantization  in an expanding
universe without imposing any subsidiary condition.

\section{Quantization without the Lorenz condition}

In the previous section it has been shown that 
although the Lorenz gauge-fixing conditions can be formally imposed in the covariant
formalism, this cannot be done in a straightforward way.   
These difficulties could be suggesting  some more fundamental obstacle 
in the formulation of an electromagnetic gauge invariant theory in an expanding universe.  
As a matter of fact, 
electromagnetic models which break gauge invariance on cosmological 
scales have been widely considered in the context of generation
of primordial magnetic fields (see, for instance, \cite{Turner}).

Let us then explore the possibility that the fundamental 
theory of electromagentism is not given by the gauge invariant 
 action (\ref{action}), but by the gauge non-invariant action (\ref{actionF}). 
Abandoning gauge invariance could, in principle, pose important problems for
the viability of the theory, namely:
\begin{itemize}
\item Modification of classical Maxwell's equations
\item Electric charge non-conservation
\item New unobserved photon polarizations
\item Negative norm (energy) states
\item Conflicts with QED phenomenology
\end{itemize}
However, as we will show in the following, none of these 
problems is actually present for the theory in  (\ref{actionF}). 
    
Since  the fundamental electromagnetic 
theory is assumed to be non-invariant under arbitrary gauge transformations, then
there is no need to impose the Lorenz constraint in the 
quantization procedure. Therefore,  having removed one constraint, 
the theory  contains one additional degree of freedom. Thus, 
the general solution for the modified equations (\ref{EMeqexp})
can be written as:
\begin{eqnarray}
\A_\mu=\A_\mu^{(1)}+\A_\mu^{ (2)}+\A_\mu^{(s)}+\partial_\mu \theta
\end{eqnarray}
where $\A_\mu^{(i)}$ with $i=1,2$ are the two transverse modes of
the massless photon, $\A_\mu^{(s)}$ is the new scalar state, which
is the mode that would have been eliminated if we had imposed the
Lorenz condition and, finally, $\partial_\mu \theta$ is a purely
residual gauge mode, which can be eliminated by means of a
residual gauge transformation in the asymptotically free regions, 
in a completely analogous way to the elimination of the $A_0$
component in the Coulomb quantization.  The fact that
Maxwell's electromagnetism could contain an additional scalar
mode decoupled from electromagnetic currents, but with 
 non-vanishing  gravitational interactions, was already noticed 
in a different context in \cite{Deser}. 

In order to quantize the free theory, we perform the mode
expansion of the  field with the corresponding creation and
annihilation operators for the {\it three} physical states:
\begin{eqnarray}
\A_{\mu}=\int d^3\vec{k}\ \sum_{\lambda=1, 2,s}\left[{\bf
a}_\lambda(k)\A_{\mu k}^{(\lambda)} +{\bf
a}_\lambda^\dagger(k)\overline{\A_{\mu k}^{(\lambda)}}\, \right]
\end{eqnarray}
where the modes are required to be orthonormal with respect to the
scalar product (see for instance \cite{Pfenning}):
\begin{eqnarray}
\left(\A^{(\lambda)}_k,\A^{(\lambda')}_{k'}\right)&=&i\int_{\Sigma}d\Sigma_\mu\left[\,
\overline{\A_{\nu k}^{(\lambda) }}\;\Pi^{(\lambda')\mu\nu}_{k'}-
\overline{\Pi^{(\lambda)\mu\nu }_{k}}\;\A_{\nu k'}^{(\lambda')}\right]\nonumber\\
&=&\delta_{\lambda\lambda'}\delta^{(3)}(\vec k-\vec k'),\;\;\;\;\;
\lambda,\lambda'=1,2,s
\label{norm}
\end{eqnarray}
where $d\Sigma_\mu$ is the three-volume element of the Cauchy
hypersurfaces. In a Robertson-Walker metric in conformal time, it
reads $d\Sigma_\mu= a^4(\eta)(d^3x,0,0,0)$. The generalized
conjugate momenta are defined as:
\begin{eqnarray}
\Pi^{\mu\nu}=-(F^{\mu\nu}-\xi g^{\mu\nu}\nabla_\rho A^\rho)
\end{eqnarray}
Notice that
the three modes can be chosen to have positive normalization.
The equal-time commutation relations:
\begin{eqnarray}
\left[\A_\mu(\eta,\vec x),\A_\nu(\eta,\vec x\,')\right]=
\left[\Pi^{0\mu}(\eta,\vec x),\Pi^{0\nu}(\eta,\vec x,')\right]=0
\end{eqnarray}
and
\begin{eqnarray}
\left[\A_\mu(\eta,\vec x),\Pi^{0\nu}(\eta,\vec x\,')\right]=
i\frac{\delta_\mu^{\;\nu}}{\sqrt{g}}\delta^{(3)}(\vec x-\vec x\,')
\end{eqnarray}
can be seen to imply  the canonical commutation relations:
\begin{eqnarray}
\left[{\bf a}_\lambda(\vec{k}),{\bf a}_{\lambda'}^\dagger(\vec{k'})\right]
=\delta_{\lambda\lambda'}\delta^{(3)}(\vec{k}-\vec{k'}),\;\;\;
\lambda,\lambda'=1,2,s
\end{eqnarray}
by means of the  normalization condition in (\ref{norm}).
Notice that the sign of the commutators is positive for the
three physical states, i.e. there are no negative norm states
in the theory, which in turn implies that there are no
negative energy states as we will see below in an
explicit example.

Since  $\nabla_\mu\A^{\mu}$ evolves as a minimally coupled scalar field,
as shown in
(\ref{minimal}), on sub-Hubble scales ($\vert k\eta\vert \gg 1$),
we find that
for arbitrary background evolution,
$\vert \nabla_\mu\A^{(s)\mu}_k\vert \propto a^{-1}$, i.e.
the field is suppressed by the universe expansion, thus 
effectively recovering the Lorenz condition on small scales. Notice that this is  
a consequence of the cosmological evolution,  not being imposed 
 as a boundary condition as in the flat space-time case.

On the other hand,
 on super-Hubble scales ($\vert k\eta\vert \ll 1$),
$\vert\nabla_\mu\A^{(s)\mu}_k\vert= const.$ which, as shown in
\cite{EM}, implies that the field contributes as a cosmological
constant in (\ref{actionF}). Indeed, the energy-momentum tensor 
derived from (\ref{actionF}) reads:
\begin{eqnarray}
T_{\mu\nu}&=&-F_{\mu\alpha}F_\nu^{\;\;\alpha}
+\frac{1}{4}g_{\mu\nu}F_{\alpha\beta}F^{\alpha\beta}\nonumber\\
&+&\frac{\xi}{2}\left[g_{\mu\nu}\left[\left(\nabla_\alpha
A^\alpha\right)^2 +2A^\alpha\nabla_\alpha\left(\nabla_\beta
A^\beta\right)\right] -4A_{(\mu}\nabla_{\nu)}\left(\nabla_\alpha
A^\alpha\right)\right]\label{Tmn}
\end{eqnarray}
Notice that, for the scalar electromagnetic mode in the 
super-Hubble limit, the contributions 
involving $F_{\mu\nu}$ vanish and only the piece proportional
to $\xi$ is relevant. 
 Thus, it can be easily seen that, since in this case
$\nabla_\alpha A^\alpha=constant$, the energy-momentum tensor is just given by:
\begin{eqnarray}
T_{\mu\nu}=\frac{\xi}{2} g_{\mu\nu}(\nabla_\alpha A^\alpha)^2
\end{eqnarray}
which is the energy-momentum tensor of a cosmological constant.  
Notice that,  as seen in (\ref{minimal}), the new scalar mode is a massless free field.  
This is one of the most relevant aspects of the present model
in which, unlike existing dark energy theories based on scalar fields, 
dark energy can be generated without including any potential term
or dimensional constant.

Since, as shown above, the field amplitude
remains frozen on super-Hubble scales and starts decaying once the
mode enters the horizon in the radiation or matter eras, the
effect of the $\xi$ term in (\ref{EMeqexp}) is completely
negligible on sub-Hubble scales, since the initial amplitude
generated during inflation is very small as we will show below. 
Thus, below 1.3 AU,
which is the largest distance scale at which electromagnetism has
been tested \cite{Nieto}, the modified Maxwell's equations
(\ref{EMeqexp}) are physically indistinguishable from the flat
space-time ones (\ref{Maxwell}).


Notice that in Minkowski space-time, the  theory (\ref{actionF})
is completely equivalent to standard QED. This is so because, although
non-gauge invariant, the corresponding effective action is equivalent to the 
standard BRS invariant effective action of QED. 
Thus, the effective action for
QED obtained from (\ref{Maxwell}) by the standard gauge-fixing procedure reads:
\begin{eqnarray}
e^{iW}=\int [dA][dc][d\bar c][d\psi][d\bar\psi] e^{i\int d^4x 
\left(-\frac{1}{4}F_{\mu\nu}F^{\mu\nu}+\frac{\xi}{2}
(\partial_\mu A^\mu)^2+\eta^{\mu\nu}\partial_\mu\bar c\, \partial_\nu c
+ {\cal L}_F\right)}
\end{eqnarray}
where ${\cal L}_F$ is the Lagrangian density of charged fermions. The $\xi$
term and the ghosts field appear in the Faddeev-Popov procedure when 
selecting an element of each gauge orbit.  
However, ghosts being decoupled from the electromagnetic currents can be integrated
out in flat space-time, so that up to an irrelevant normalization constant we find:
\begin{eqnarray}
e^{iW}=\int [dA][d\psi][d\bar\psi] 
e^{i\int d^4x \left(-\frac{1}{4}F_{\mu\nu}F^{\mu\nu}+\frac{\xi}{2}
(\partial_\mu A^\mu)^2+ {\cal L}_F\right)}
\end{eqnarray}
which is nothing but the effective action coming from 
the gauge non-invariant theory (\ref{actionF}) in flat space-time, in which
no  gauge-fixing procedure is required.

To summarize, none of the above mentioned consistency problems for the theory
in (\ref{actionF}) arise, thus:
\begin{itemize}
\item Ordinary Maxwell's equations are recovered on those small scales
in which electromagnetism has been tested.
\item Electric charge is conserved since only the gauge electromagnetic sector
is modified but not the sector of charged particles which preserves its gauge symmetry.
\item The new state only couples gravitationally and evades laboratory detection.
\item The new state has positive norm (energy).
\item The effective action is completely equivalent to standard QED in the flat
space-time limit. This guarantees, not only that the standard phenomenology is
recovered, but also that no new interaction terms will appear in the renormalization
procedure. 
\end{itemize}

\section{Cosmological evolution}
Let us now consider the cosmological evolution of this new electromagnetic mode. For 
that purpose, we will consider an homogeneous electromagnetic field with 
$A_\mu=(A_0(t),\vec A(t))$. The corresponding equations motion in a flat
Robertson-Walker background in cosmological time $t$ read:
\begin{eqnarray}
\ddot{A}_0&+&3H\dot{A}_0+3\dot{H}A_0=0\label{const}
\nonumber\\
\ddot{\vec{A}}&+&H\dot{\vec{A}}=0
\end{eqnarray}
Notice that conformal time components are related to components in 
cosmological time as $\A_\mu=(aA_0,\vec{A})$.
 The effects of the high electric conductivity of the universe
can be taken into account by including the corresponding current term $J^\mu$ on the 
right hand side of Maxwell's equations, however due to the strict
electric neutrality of the universe on large scales $J^\mu=0$.

In the case in which the scale factor behaves as a simple
power law with $H=p/t$, the solutions for the above equations 
are:
\begin{eqnarray}
A_0(t)&=&A_0^+t+A_0^-t^{-3p}\nonumber\\
\vec{A}(t)&=&\vec{A}^+t^{1-p}+\vec{A}^-\label{Azsol} 
\end{eqnarray}
Thus we see that the temporal component grows faster than the spatial 
one and therefore, at late times, on cosmological scales we can ignore the spatial
contribution and the new scalar state is essentially given by the
electric potential. Notice also that from the temporal equation we get:
\begin{eqnarray}
\frac{d}{dt}(\nabla_\mu A^\mu)=\frac{d}{dt}(\dot A_0+3HA_0)=0
\end{eqnarray}
irrespective of the background evolution, i.e. on large scales the
scalar mode $\nabla_\mu A^\mu$ is strictly constant as commented before.
We can also compute the contributions from the temporal and spatial components
to the energy density from (\ref{Tmn}), thus we get:
\begin{eqnarray}
\rho_{A_0}&=&\frac{\lambda}{2}\left(\dot A_0+3HA_0\right)^2=\,const.\nonumber\\
\rho_{\vec{A}}&=&\frac{1}{2a^2}\left(\dot{\vec{A}}\right)^2\propto\,a^{-4} 
\end{eqnarray}
We see that also, as commented before, the temporal component behaves as 
a cosmological constant, whereas the contibution from the spatial part decays
as radiation and therefore does not affect the universe isotropy on large
scales.
\begin{figure}[h]
  \includegraphics[height=.3\textheight]{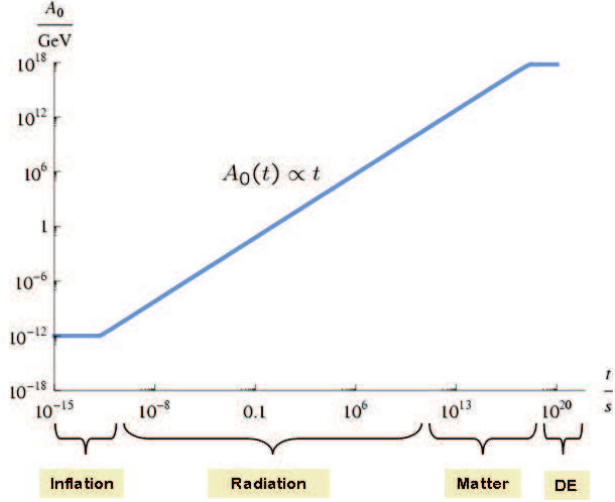}
  \caption{Cosmological evolution of the temporal component
from electroweak-scale inflation until present}
\end{figure}

In Fig.2 we show the cosmological evolution of the electric potential $A_0$.
We see that the field is constant during inflation, it grows linearly in
time in the matter and radiation eras and becomes also constant when 
the eletromagnetic dark energy starts dominating. Notice that the present
value of $A_0$ will be determined by the initial value of the field
generated during inflation from quantum fluctuations. This amplitude can 
be explicitly calculated as shown in next section.

\section{Quantum fluctuations during inflation}

Let us consider the  quantization in an inflationary de Sitter space-time with
$a(\eta)=-1/(H_I\eta)$. Here we will take $\xi=1/3$, although similar
results can be obtained for any $\xi>0$.  
The explicit solution for the normalized
scalar state is \cite{EM}:
\begin{eqnarray}
\A_{0k}^{(s)}&=&-\frac{1}{(2\pi)^{3/2}}\frac{i}{\sqrt{2k}}
\left\{k\eta e^{-ik\eta}\right.\nonumber\\
&+&\left.\frac{1}{k\eta}\left[\frac{1}{2}(1+ ik\eta)
e^{-ik\eta}-k^2\eta^2e^{ik\eta}E_1(2ik\eta)\right]\right\}e^{i\vec k \vec x}\nonumber\\
\nonumber\\
\A_{\parallel k}^{(s)}&=&\frac{1}{(2\pi)^{3/2}}\frac{1}{\sqrt{2k}}
\left\{(1+ik\eta)e^{-ik\eta}\right.\nonumber\\
&-&\left.\left[\frac{3}{2}e^{-ik\eta}+(1-ik\eta)e^{ik\eta}
E_1(2ik\eta)\right]\right\}e^{i\vec k \vec x}\label{scalar}
\end{eqnarray}
 where $E_1(x)=\int_1^\infty e^{-tx}/tdt$ is the exponential
integral function. Using this solution, we  find:
\begin{eqnarray}
\nabla_\mu\A^{(s)\mu}_k=-\frac{a^{-2}(\eta)}{(2\pi)^{3/2}}
\frac{ik}{\sqrt{2k}}\frac{3}{2}\frac{(1+ik\eta)}
{k^2\eta^2}e^{-ik\eta+i\vec k \vec x}
\end{eqnarray}
so that the field is suppressed in the sub-Hubble limit
as  $\nabla_\mu\A^{(s)\mu}_k\sim \Od((k\eta)^{-2})$.

On the other hand, from the energy density given by $\rho_A= T^0_{\;\;0}$,
we obtain in the sub-Hubble limit the corresponding Hamiltonian, which is
given by:
\begin{eqnarray}
H=\frac{1}{2}\int \frac{d^3\vec
k}{a^4(\eta)}k\sum_{\lambda=1,2,s}\left[ {\bf
a}_{\lambda}^{\dagger}(\vec k){\bf a}_{\lambda}(\vec k) +{\bf
a}_{\lambda}(\vec k){\bf a}_{\lambda}^{\dagger}(\vec k) \right].
\end{eqnarray}
We see that the theory does not contain negative energy states (ghosts).

Finally, from (\ref{scalar}) it is possible to obtain the
dispersion of the effective cosmological constant during
inflation:
\begin{eqnarray}
\langle 0\vert(\nabla_\mu\A^{\mu})^2\vert 0 \rangle=\int\frac{dk}{k}P_A(k)
\end{eqnarray}
with $P_A(k)=4\pi k^3\vert\nabla_\mu\A^{(s)\mu}_k\vert^2 $. In the
super-Hubble limit, we obtain for the power-spectrum:
\begin{eqnarray}
P_A(k)=\frac{9H_I^4}{16\pi^2},
\end{eqnarray}
in agreement with \cite{EM}. Notice that this
result implies that $\rho_A\sim (H_I)^4$. The measured value of
the cosmological constant then requires $H_I\sim 10^{-3}$ eV,
which corresponds to an inflationary scale  $M_I\sim 1$ TeV.
Thus we see that the cosmological constant scale can be naturally
explained in terms of physics at the electroweak scale.

\section{Perturbations and consistency}

Despite the fact that the background evolution in the present case
is the same as
in $\Lambda$CDM, the evolution of metric perturbations could
be different, thus offering an observational way of discriminating
between the two models. With this purpose, we have calculated the evolution of metric,
matter density and
electromagnetic perturbations \cite{BKMM}. The propagation speeds
of scalar, vector and tensor perturbations are found
to be real and equal to the speed of light, so that the theory is
classically stable. We have also shown before that  the theory does
not contain ghosts and it is therefore stable at the quantum level.
On the other hand, using the explicit expressions in \cite{Will} for
the  vector-tensor theory of gravity corresponding to the action in
(\ref{actionF}),  it is
possible to see that all the parametrized post-Newtonian (PPN) parameters
agree with those of General Relativity,  i.e. the theory is compatible
with all the local gravity constraints for any value
of the homogeneous background electromagnetic field \cite{EM,VT}.

Concerning the evolution of
scalar perturbations, 
we find
that  the only relevant deviations with respect to $\Lambda$CDM
appear on large scales $k\sim H_0$ and that
they depend on the primordial
spectrum of electromagnetic fluctuations. However,
the effects on the CMB temperature and matter power spectra 
are compatible with observations except for very large primordial
fluctuations.
 In addition, the different evolution of the scalar potential $\Phi_k$
with respect to the $\Lambda$CDM model gives rise to
a possible discriminating  contribution to the
late-time integrated Sachs-Wolfe effect \cite{Turok}.

\section{Conclusions}
In this work we have presented a consistent quantization
procedure for the electromagnetic interaction with three polarization
states. The energy density of the new scalar mode on cosmological scales 
is shown to behave as an effective cosmological constant, whose value is determined 
by the amplitude of  quantum fluctuations generated during inflation. 
As a matter of fact, the measured value
of the cosmological constant is naturally explained provided inflation
took place at the electroweak scale. The model is free from classical 
or quantum instablities and is consistent with all the local gravity
constraints. On the other hand, it is also compatible with observations
from CMB and large scale structure and contains the same number of free
parameters as $\Lambda$CDM. Unlike dark energy models
based on scalar fields, acceleration in this model arises from the kinetic
term of the new electromagnetic mode, without the introduction of unnatural
dimensional parameters or potential terms. The results presented in this work
show that, unlike previous proposals,  the nature of dark 
energy can be established without resorting to new physics.

\vspace{0.2cm}

{\em Acknowledgments:}
 This work has been  supported by
Ministerio de Ciencia e Innovaci\'on (Spain) project numbers
FIS 2008-01323 and FPA
2008-00592, UCM-Santander PR34/07-15875, CAM/UCM 910309 and
MEC grant BES-2006-12059.

\end{document}